\documentclass[reprint,pra,aps,showpacs,amsmath,amssymb]{revtex4-1}
\usepackage{graphicx}

\begin{document}

\title{Experimental implementation of a fully controllable depolarizing quantum operation}

\author{Youn-Chang \surname{Jeong}}\email{w31400@gmail.com}

\author{Jong-Chan \surname{Lee}}

\author{Yoon-Ho \surname{Kim}}\email{yoonho72@gmail.com}
\affiliation{Department of Physics, Pohang University of Science and Technology (POSTECH), Pohang, 790-784, Korea}

\date{\today}

\begin{abstract}
The depolarizing quantum operation plays an important role in studying the quantum noise effect and implementing general quantum operations. In this work, we report a scheme which implements  a fully controllable input-state independent depolarizing quantum operation for a photonic polarization qubit.
\end{abstract}



\maketitle

One of the main challenges in experimental quantum information is to deal with decoherence. To preserve a qubit state, it is often necessary to isolate the qubit from the environment but to be able to do information processing tasks, interactions with an external (classical or quantum) system is necessary. More often than not, unwanted interactions with the environment leave the qubit in a different quantum state or even cause the qubit to lose coherence. Such unwanted quantum state transformation can be described by the quantum process due to a noisy quantum channel, in other words,  a noisy quantum process, which can be understood with a few basic single-qubit noise operations including the bit-flip, phase-flip, depolarization, and amplitude damping \cite{Nielsen2000}.

Understanding and being able to implement each of the basic quantum noise processes are important in theoretical and experimental quantum information science.  First, it would allow us to simulate and quantify  quantum noise processes. Second, the bit-flip and the phase-flip operations are essential for quantum error correction protocols.  Third, the depolarization and the amplitude damping operations can describe decoherence.  

The depolarization operation is of particular interest because, in addition to being relevant to a number of practical quantum communication and computation scenarios \cite{Bruss2000,Ricci2004,Dawson2006,Cafaro2010,Jeong2011}, it is also an essential operation for optimally approximating non-physical quantum operations \cite{Lim2011a,Lim2011b} and for generating exotic quantum states including Werner states \cite{Barbieri2004} and bound entangled states \cite{Lavoie2010}. Clearly, from the experimental point of view, it is important to develop a method to achieve a fully controllable input-independent depolarization quantum operation. 

As such, there have been many reports on experimental implementation of a depolarization quantum channel. In Ref.~\cite{Puentes2005}, optical scattering media were used to achieve the depolarization quantum channel but the scheme naturally induces the spread in the photon momenta and it is difficult to control the degree of depolarization in this scheme.  A controllable depolarization channel was demonstrated in Ref.~\cite{Puentes2006,Shaham2011}  but the output was dependent on the input polarization and the scheme also introduced the momentum spread for the photon. Input-independent depolarizing quantum operation was demonstrated in Ref.~\cite{Ricci2004,Karpinski2008}, but here the depolarizing operation was achieved by time averaging of many `fast' operations, i.e., incoherent sum of many different pure quantum states. 
 
In this paper, we report an experimental implementation of a fully-controllable  depolarization quantum operation for a photonic polarization qubit. The scheme is completely input-state independent so that it is possible to introduce any desired degree of depolarization regardless of the state of the input qubit. Furthermore, our scheme does not rely on time- or spatial-averaging so that neither the measurement duration nor the measurement area affect the output quantum state. In other words, our scheme achieves a truly observer-independent depolarizing quantum operation.

\begin{figure}[b]
\includegraphics[width=3in]{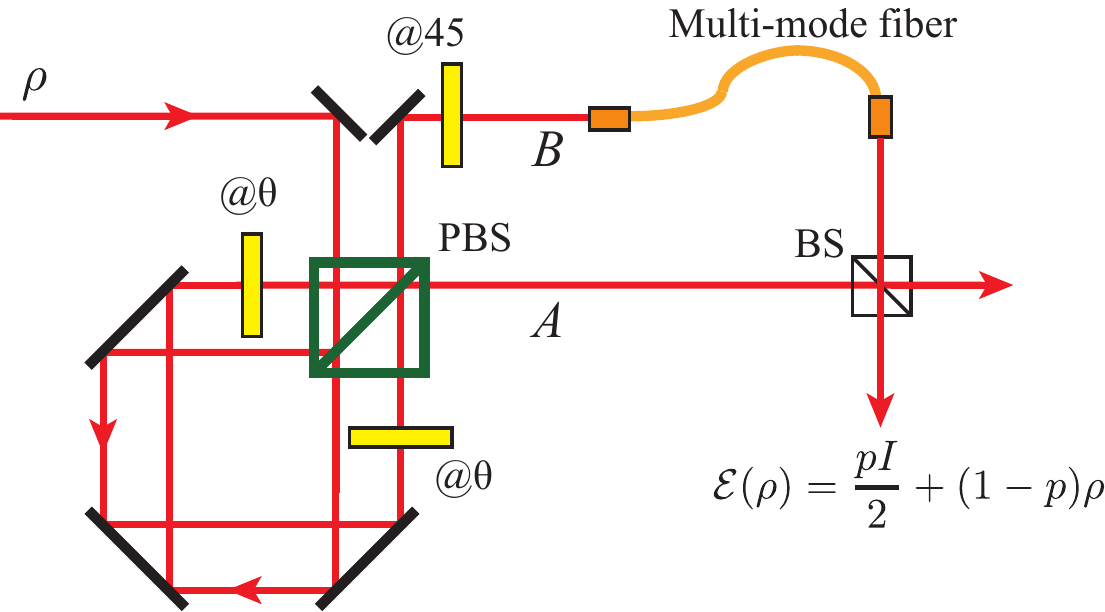}
\caption{Experimental schematic for a fully-controllable input-independent depolarizing quantum operation for a photonic polarization qubit $\rho$.
}\label{scheme}
\end{figure}

The depolarizing quantum operation is described as
\begin{equation}\label{eq:depolarizing}
\mathcal{E}(\rho) = \frac{p I}{2}+(1-p)\rho,
\end{equation}
where $p$ is the degree of decoherence ($0\leq p\leq 1$), $\rho$
is the input state of the photonic polarization qubit, and $I$ is the two-dimensional identity matrix.  We implement the depolarizing operation $\mathcal{E}(\rho)$ by using a modified displaced Sagnac interferometer setup as shown in Fig.~\ref{scheme}. The displaced Sagnac interferometer consists of a polarizing beam splitter (PBS) and two half-wave plates that can be oriented at an angle $\theta$ (@$\theta$ in Fig.~\ref{scheme}). Another half-wave plate fixed at 45$^\circ$  at the output mode $B$ makes the polarization state of the two output modes $A$ and $B$ identical.

\begin{figure}[t]
\includegraphics[width=3in]{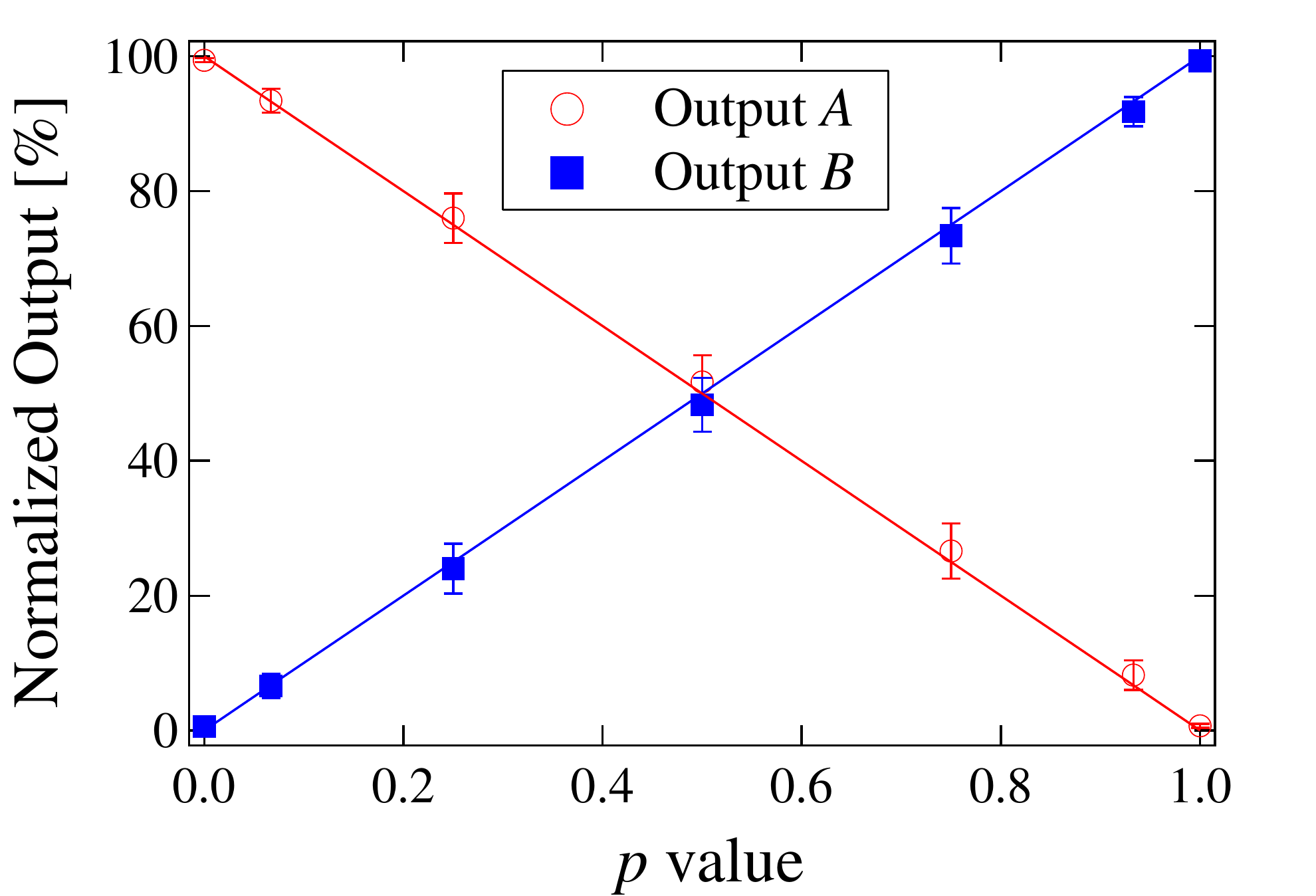}
\caption{Data showing the linear splitting ratio $A:B = 1-p:p$ where $p = \sin^2(2\theta)$ for the displaced Sagnac. Each data point represents the averaged normalized output of six input polarization states, $|H\rangle$, $|V\rangle$, $|D\rangle$, $|A\rangle$, $|R\rangle$, and $|L\rangle$. The normalized output of each polarization state appears almost identical to the averaged result shown here. 
}\label{data}
\end{figure}

It is not difficult to see that the displaced Sagnac interferometer in Fig.~\ref{scheme} acts as a continuously variable non-polarizing beam splitter in which the output ratio $A:B=1-p:p$ can be linearly varied by setting the angle $\theta$ of the two half-wave plates. Note that  the splitting parameter $p = \sin^{2}(2 \theta)$. This is confirmed in the experiment as shown in Fig.~\ref{data}. In the experiment, we prepared the single-photon polarization qubit $\rho$ by using the heralded single-photon state generated from spontaneous parametric down-conversion (SPDC) in a 3 mm type-II BBO crystal.  The wavelengths of the pump and that of the SPDC photon are 405 nm and 810 nm. The input polarization qubit was prepared in a pure state $\rho = |\psi\rangle\langle\psi|$ by using a half-wave and a quarter-wave plate.  We then measured the count rates at the two single-photon detectors placed at the output ports $A$ and $B$ and the data are shown in Fig.~\ref{data}. As expected, the data show the linear splitting ratio between two outputs $A$ and $B$ with the splitting parameter $p$. In Fig.~\ref{data}, we plot the averaged normalized outputs $A$ and $B$ for six input polarization states,  $|H\rangle$ (horizontal), $|V\rangle$ (vertical), $|D\rangle \equiv (|H\rangle + |V\rangle)/\sqrt{2}$, $|A\rangle\equiv (|H\rangle - |V\rangle)/\sqrt{2}$, $|R\rangle \equiv (|H\rangle - i|V\rangle)/\sqrt{2}$, and $|L\rangle \equiv (|H\rangle +i|V\rangle)/\sqrt{2}$. The normalized outputs $A$ and $B$ for each polarization state look almost identical to the averaged result shown in Fig.~\ref{data}. We also point out that the linearity and the splitting ratio are very stable over time due to the Sagnac geometry. 

Previously reported continuously variable non-polarizing beam splitters by using a prism pair \cite{Harrick1963}, a sapphire disk \cite{Broer1981}, and a phase grating \cite{Zhong2004} were all input-polarization dependent. However, our scheme based on the displaced Sagnac interferometer is completely input-polarization independent as demonstrated in Fig.~\ref{data}. 

To be sure that the displaced Sagnac splits the input probability amplitude into two spatial modes $A$ and $B$ without affecting the quantum state $\rho$, we have also performed the quantum state tomography of the input qubit as well as the output qubit in modes $A$ and $B$ \cite{James2001}. We observed that the fidelity between the input and the output quantum states is better than $0.982 \pm 0.003$ for all  polarization qubit states we have tested. Thus, the displaced Sagnac acts as a nearly ideal identity operation for the polarization qubit except that it diverts the amplitude into two different spatial modes $A$ and $B$.

\begin{figure}[t]
\includegraphics[width=2.6in]{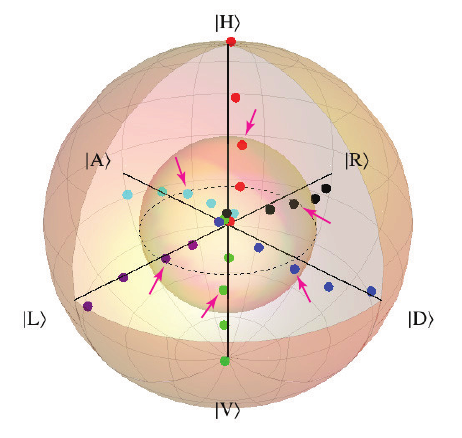}
\caption{Experimental data. As $p$ is increased, the qubit states become more mixed, hence moving toward the center of the Bloch sphere. The outer sphere represents $p=0$ (pure states) and the inner sphere represents
$p=0.5$. The arrows point the data points (i.e., qubit states) on the inner sphere.}\label{depol}
\end{figure}

Equation (1) clearly states that, to implement the depolarizing quantum operation $\mathcal{E}(\rho)$, it is necessary to achieve a mixture of the input quantum state $\rho$ and the unpolarized state $I/2$ with the weighting factors $1-p$ and $p$, respectively. Note now that the qubit states found at the outputs $A$ and $B$  are identical to the input  $\rho$ but with the probability amplitudes $1-p$ and $p$, respectively. Thus, we first couple the output mode $B$ into a 2-m long multi-mode fiber to transform a polarized input to a completely unpolarized state. We then combine the output of the multi-mode fiber (after collimation) and that of $A$ at a beam splitter (BS). The 2-m long multi-mode fiber ensures that the beam combination at BS is a completely incoherent process since the path length difference is orders of magnitude larger than the single-photon coherence time, which is on the order of  hundreds of femtoseconds.  Therefore, the quantum state found at the two outputs of the BS is described precisely as $pI/2 + (1-p)\rho$, indicating that the the input state $\rho$ has gone through the depolarizing quantum operation $\mathcal{E}(\rho)$  in eq.~(1). 

To demonstrate that the outputs of the BS indeed correspond to the quantum state after the depolarizing quantum operation, we performed quantum state tomography on the output states for six different input qubit states. The experimental data are shown in Fig.~\ref{depol}.  It is clear that, by increasing $p$, the qubit states become more mixed, moving toward the center of the Bloch sphere. It is important to note that the depolarizing quantum operation is an isotropic operation so that the output state purity $Tr[\rho^2]$ does not depend on the input state: it only depends on the choice $p$. As such, the depolarizing quantum operation should only shrink the size of the Bloch sphere, rather than making it asymmetric. This feature is well demonstrated in Fig.~\ref{depol}: all the data points corresponding to the depolarizing quantum operation of the same $p$, regardless of the input state, should reside on the same sphere. In Fig.~\ref{depol}, the arrows represent the qubit states after  the depolarizing quantum operation $\mathcal{E}(\rho)$ for $p=0.5$ and they all lie on the inner sphere representing all qubit states undergone $\mathcal{E}(\rho)$ with $p=0.5$.  We also note that changing $p$ is quite easy in our setup as it requires only the rotation of the waveplates due to the relation $p=\sin^2(2\theta)$.

\begin{figure}[t]
\includegraphics[width=3.2in]{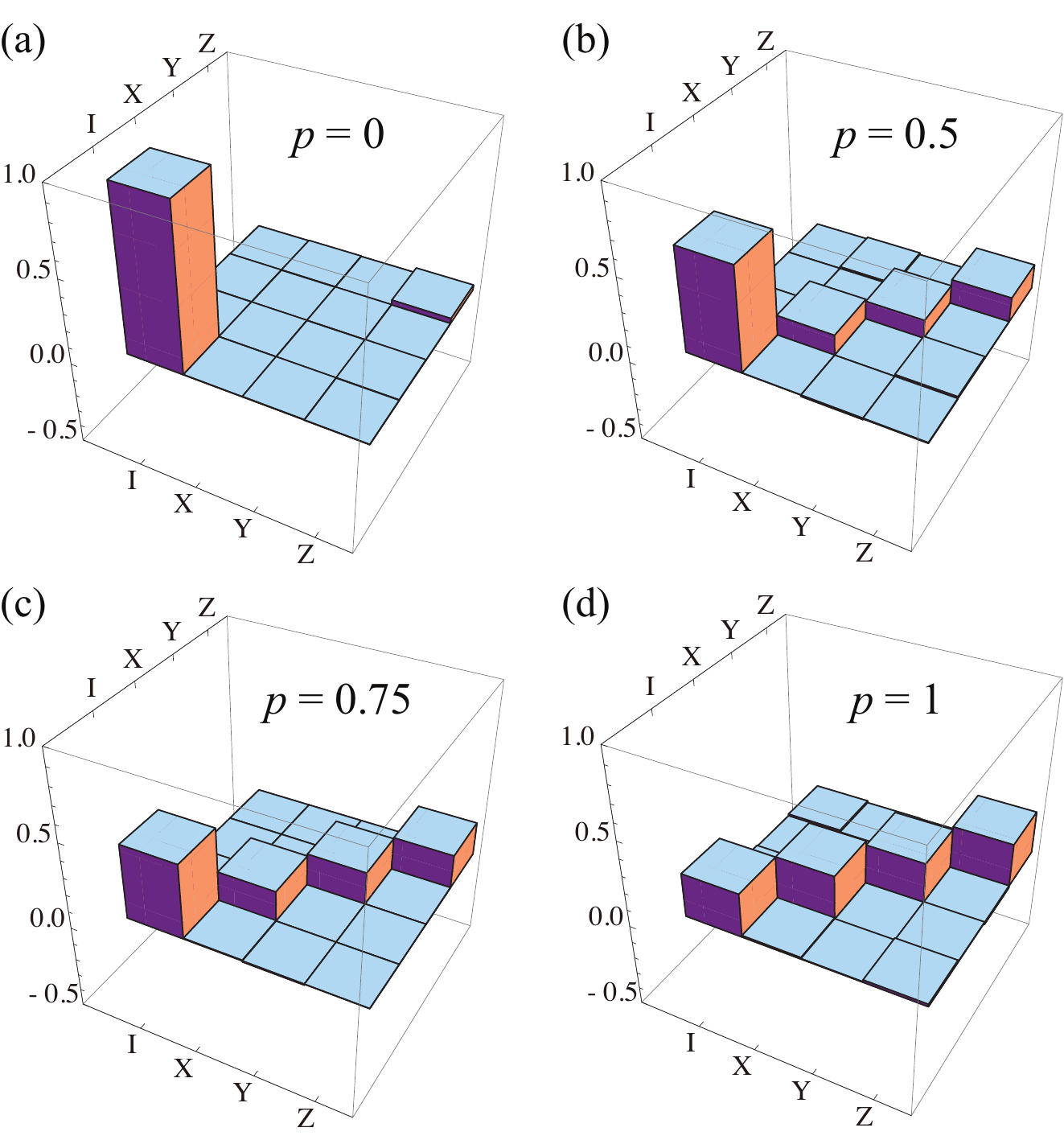}
\caption{Real part of $\chi$ matrices of the depolarizing quantum
channel. $I$ refers to the two dimensional identity matrix. $X$, $Y$, and $Z$ correspond to Pauli operators.
 The process fidelities (in comparison to the ideal operation) are (a)  0.966,    
(b) 0.994, (c) 0.998, and (d) 0.997. The imaginary part of the $\chi$ matrices are almost zero, hence not shown.
}\label{kai}
\end{figure}

It is known that a quantum channel  that implements a particular quantum operation can be fully characterized by performing quantum process tomography  \cite{Fiurasek2001}. We have carried out quantum process tomography for the depolarizing quantum operation with various $p$ and the resulting $\chi$ matrices are shown in Fig.~\ref{kai}. For a single-qubit operation as this one, it is often best to use the Pauli-basis for decomposing the quantum process as done in the figure. Clearly, when $p=0$, the quantum process corresponds to an identity operation as it should be, see Fig.~\ref{kai}(a). As $p$ is increased, contributions from the Pauli operations rise, see Fig.~\ref{kai}(b) and \ref{kai}(c), and when $p=1$, it is clear that the output state will be a fully mixed state regardless of the input state, see Fig.~\ref{kai}(d). The high process fidelities for various $p$ values indicate the robustness of our setup to faithfully implement the fully controllable depolarizing quantum operation.

In summary, we have reported an experimental realization of a fully controllable depolarizing quantum operation for a single-photon polarization qubit. Our scheme not only allows continuous adjustment of the degree of depolarization but also is independent of the input quantum state, as demonstrated with quantum state tomography and quantum process tomography. A versatile depolarizing quantum channel like the one reported in this paper should find applications in many areas of photonic quantum information research, including generating exotic quantum states,  studying the quantum noise processes, approximating non-physical quantum operations, etc.

This work was supported in part by the National Research Foundation of Korea (2009-0070668 and 2011-0021452).

\end{document}